\documentclass[12pt,letterpaper]{article}

\usepackage{amsmath,amsfonts,amssymb,srcltx,amsthm}
\usepackage{graphicx}
\usepackage{caption}
\usepackage{subcaption}

\usepackage{url}
\newcommand{\eb}{\pmb{e}}

\usepackage[margin=1in]{geometry}

\usepackage[round]{natbib}
\usepackage{setspace}
\usepackage{caption}

\newtheorem{theorem}{Theorem}[section]
\newtheorem{lemma}[theorem]{Lemma}

\title{Generalized Hultman Numbers and Cycle Structures of Breakpoint Graphs}
\author{
Nikita Alexeev,\thanks{The George Washington University, Washington, DC, USA}~\footnotemark[3]~
Anna Pologova,\thanks{St. Petersburg State University, St. Petersburg, Russia}~
and~
Max A. Alekseyev\footnotemark[1]
}

\date{}

\begin{document}

\maketitle

\renewcommand*{\thefootnote}{\fnsymbol{footnote}}
\footnotetext[3]{Corresponding author. Email: \texttt{nikita\_alexeev@gwu.edu}}
\renewcommand*{\thefootnote}{\arabic{footnote}}

 \begin{abstract}
  Genome rearrangements can be modeled as $k$-breaks, which break a genome at $k$ positions and glue the resulting fragments in a new order.
 In particular, reversals, translocations, fusions, and fissions are modeled as $2$-breaks, and transpositions are modeled as $3$-breaks.
 While $k$-break rearrangements for $k>3$ have not been observed in evolution,
 they are used in cancer genomics to model chromothripsis, a catastrophic event of multiple breakages happening simultaneously in a genome.
 It is known that the $k$-break distance between two genomes (i.e., the minimum number of $k$-breaks required to transform one genome into the other)
 can be computed in terms of cycle lengths in the breakpoint graph of these genomes.

 In the current work, we address the combinatorial problem of enumerating genomes at a given $k$-break distance from a fixed unichromosomal genome. More generally, we enumerate genome pairs, whose breakpoint graph has a given distribution of cycle lengths.
 We further show how our enumeration can be used for uniform sampling of random genomes at a given $k$-break distance, and describe its connection to various combinatorial objects such as Bell polynomials.
\end{abstract}

\section{Introduction}

Genome rearrangements are evolutionary events that change gene order along the genome.
The genome rearrangements can be modeled as \emph{$k$-breaks}~\citep{alekseyev2008},
which break a genome at $k$ positions and glue the resulting fragments in a new order.
While most frequent genome rearrangements such as \emph{reversals} (which flip segments of a chromosome),
\emph{translocations} (which exchange segments of two chromosomes),
\emph{fusions} (which merge two chromosomes into one), and \emph{fissions} (which split a single chromosome into two)
can be modeled as $2$-breaks (also called \emph{Double-Cut-and-Join} or \emph{DCJ} in \citealt{yancopoulos2005}),
more complex and rare genome rearrangements such as transpositions are modeled as $3$-breaks.
While $k$-break rearrangements for $k>3$ have not been observed in evolution, they are used in cancer genomics to model \emph{chromothripsis},
a catastrophic event of multiple breakages happening simultaneously in the genome~\citep{chromo2011,weinreb2014}.

The \emph{$k$-break distance} between two genomes is defined as the minimum number of $k$-breaks required to transform one genome into the other.
The $2$-break (DCJ) distance is often used in phylogenomic studies to estimate the evolutionary remoteness of genomes.
The $k$-break distance between two genomes can be expressed in terms of cycles in the breakpoint graph of these genomes.
Namely, while the $2$-break distance depends only on the number of cycles in this graph, the $k$-break distance in general
depends on the distribution of the cycle lengths~\citep{alekseyev2008}.

In the current work, we address the combinatorial enumeration of genomes at a given $k$-break distance from a fixed unichromosomal genome.
More generally, for a fixed unichromosomal genome $P$, we enumerate all genomes $Q$ such that the breakpoint graph of $P$ and $Q$ has a given distribution of cycle lengths.
We consider various flavors of this problem, where genes may be arbitrarily oriented
or co-oriented along the genomes,\footnote{The case of unoriented genes is presumably much harder. For example, computing the reversal distance between unichromosomal genomes with unoriented genes is known to be NP-hard~\citep{Caprara1997}.}
while the genomes $Q$ may be unichromosomal or multichromosomal.
In the multichromosomal case we restrict genomes to contain only circular chromosomes,
while in the unichromosomal case we consider both circular and linear genomes.

Previous studies are mostly concerned with $2$-break distances between unichromosomal genomes.
In particular, unichromosomal genomes with co-oriented genes can be interpreted as permutations,
and the number of permutations at a given $2$-break distance from the identity permutation is given by Hultman numbers~\citep{hultman1999}.
\citet{doignon2007} gave a closed formula for Hultman numbers, \citet{bona2009} proved a relation between Hultman numbers and Stirling numbers of the first kind.
The case of $2$-break distances between genomes with arbitrarily oriented genes was solved by \citet{signed2013}.
The asymptotic distribution of $2$-break distances was proved to be normal by \citet{alexeev2014}.
The analog of Hultman numbers for multichromosomal circular genomes was recently studied by \citet{multi2014}.
The current work generalizes all these results.

\section{Background}

We start our analysis with (multichromosomal) circular genomes and later extend it to unichromosomal linear genomes.

\begin{figure}[!t]
 \begin{center}
 \includegraphics[width=\textwidth]{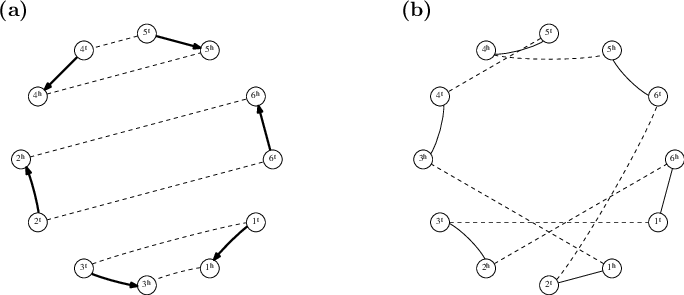}
  \caption{For genomes $P=(1,2,3,4,5,6)$ and $Q=(1,-3)(2,-6)(4,-5)$, \textbf{(a)} the genome graph of $Q$; \textbf{(b)} the breakpoint graph $G(P,Q)$,
  where the adjacency edges of $P$ and $Q$ are colored black (solid) and gray (dashed), respectively. 
  The graph $G(P,Q)$ consists of one 2-cycle and one 4-cycle.}
  \label{fig:gen_bpg}
 \end{center}
 \end{figure}

We represent a circular genome consisting of genes $\{1,2,\dots,n\}$ as a \emph{genome graph}. 
This graph contains $2n$ vertices: for each gene $i \in \{1,2,\dots,n\}$, there are the \emph{tail} and \emph{head} vertices $i^t$ and $i^h$,
The graph has $n$ directed \emph{gene edges} of the form $(i^t,i^h)$ encoding $n$ genes, 
and $n$ undirected \emph{adjacency edges} connecting neighboring head/tail vertices of adjacent genes (Fig.~\ref{fig:gen_bpg}a). 
We remark that for the genomes with co-oriented genes all adjacency edges connect the head of one gene with the tail of another. 

Let $P$ and $Q$ be a pair of circular genomes on the same genes $\{1,2,\dots,n\}$.
We assume that in their genome graphs the adjacency edges of $P$ are colored black and the adjacency edges of $Q$ 
are colored gray.
The \emph{breakpoint graph} $G(P, Q)$ is defined on the set of vertices $\{i^t,i^h\mid i=1,\dots,n\}$ 
with black and gray edges inherited from genome graphs of $P$ and $Q$ (Fig.~\ref{fig:gen_bpg}b). 
Since each vertex in $G(P,Q)$ has degree $2$, the black and gray edges form a collection of alternating black-gray cycles.
We say that a black-gray cycle is an \emph{$\ell$-cycle} if it is composed of $\ell$ black and $\ell$ gray edges. 
Let $c_{\ell}(P,Q)$ be the number of $\ell$-cycles in $G(P,Q)$. Then the total number of black edges in $G(P,Q)$ equals
$$\sum_{\ell\geq 1} \ell\cdot c_\ell(P,Q) = n \, .$$ 

\begin{figure}[!t]
 \begin{center}
 \includegraphics[width=\textwidth]{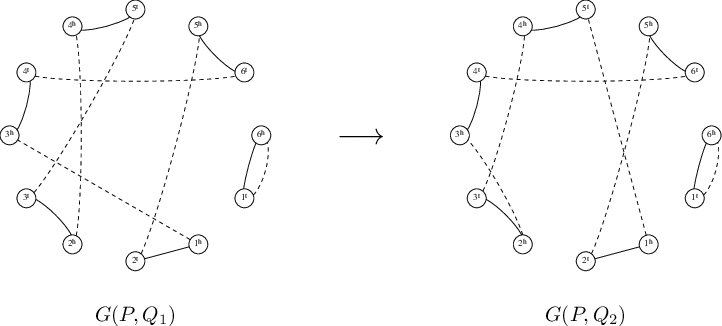}
  \caption{A 3-break transforming genome $Q_1 = (1,-3,5,2,-4,6)$ into genome $Q_2 = (1,5,2,-3,-4,6)$ corresponds to a transformation of the breakpoint graph $G(P,Q_1)$ into $G(P,Q_2)$ by
  replacing the gray edges $\{1^h,3^h\}$, $\{2^h, 4^h\}$, and $\{3^t,5^t\}$ with the gray edges $\{1^h,5^t\}$, $\{2^h,3^h\}$, and $\{3^t,4^h\}$.}
  \label{fig:kbreak}
 \end{center}
 \end{figure}

A $k$-break in genome $Q$ corresponds to an operation in its genome graph and the breakpoint graph $G(P,Q)$. Namely, a $k$-break 
replaces any $k$-tuple of gray edges with another $k$-tuple of gray edges forming a matching on the same set of $2k$ vertices (Fig.~\ref{fig:kbreak}). 
A transformation of genome $Q$ into genome $P$ with $k$-breaks can therefore be viewed as a transformation of the breakpoint graph $G(P,Q)$ 
into the breakpoint graph $G(P,P)$ with $k$-breaks on gray edges.
The \emph{$k$-break distance} $d_k(P,Q)$ between genomes $P$ and $Q$ is the minimum number of $k$-breaks in such a transformation.

The $2$-break distance between genomes $P$ and $Q$ is given by the following formula~\citep{yancopoulos2005}:
\begin{equation}
d_2(P, Q) = n-c(P, Q) \; ,
\label{eq:d2}
\end{equation}
where $c(P, Q)=\sum_{\ell\geq 1} c_{\ell}(P,Q)$ is the total number of cycles in $G(P, Q)$.
Formulae for the $k$-break distance for $k>2$ are more sophisticated. 
In particular, $d_3(P,Q)$ and $d_4(P,Q)$ are given by the following formulae~\citep{alekseyev2008}:
\begin{equation}
d_3(P,Q) = \frac{n-c^{2,1}(P,Q)}{2} \, ,\label{eq:d3}\\
\end{equation}
\begin{equation}
d_4(P,Q) = \left\lceil\frac{n-c^{3,1}(P,Q)-\lfloor c^{3,2}(P,Q)/2\rfloor}{3}\right\rceil \, , \label{eq:d4}
\end{equation}
where 
$$c^{m,i}(P,Q)=\sum_{\ell \equiv i \pmod{m}} c_\ell(P,Q) \, .$$

For a fixed unichromosomal genome $P$ with $n$ genes and a given vector $(c_1,c_2,c_3, \dots,c_n)$ of nonnegative integers
such that $\sum_{\ell =1}^n \ell\cdot c_\ell=n$, we will compute the number of genomes $Q$ such that $G(P,Q)$ consists of $c_\ell$ $\ell$-cycles (for each $\ell \in \{1,2,\dots\}$). 
As an application, this enumeration will allow us to find the distribution of $k$-break distances 
from various genomes $Q$ to a fixed genome $P$ for any $k\geq 2$.

\section{Genomes With A Fixed Breakpoint Graph}

Let $\mathbf{c} = (c_1,c_2,c_3, \dots)$ be a sequence of nonnegative integers with a finite number of nonzero (i.e., strictly positive) terms.
Then $L(\mathbf{c}) = \sum_{\ell\geq 1} \ell\cdot c_\ell$ is a finite integer.
We say that a breakpoint graph has \emph{cycle structure} $\mathbf{c}$ if for every positive integer $\ell$, the number of $\ell$-cycles in this graph equals $c_\ell$.

Let $P$ be a fixed unichromosomal genome with $n$ genes and $\mathcal{Q}_n(h;\mathbf{c})$ 
be the set of $h$-chromosomal genomes $Q$ on 
the same $n$ genes such that $G(P,Q)$ has cycle structure $\mathbf{c}$.\footnote{We remark that the genome $P$ essentially corresponds to a cyclic order on the genes of genomes $Q$. Hence, $\mathcal{Q}_n(h;\mathbf{c})$ is well-defined as soon as we are given a cyclic order on the genes. Without loss of generality, we may assume that the genes are labeled by numbers from $1$ to $n$ (up to a cyclic rotation).} 
Let $M_n(h;\mathbf{c})$ be the cardinality of $\mathcal{Q}_n(h;\mathbf{c})$, i.e., $M_n(h;\mathbf{c}) = |\mathcal{Q}_n(h;\mathbf{c})|$. Clearly, we have $M_n(h;\mathbf{c}) = 0$ unless $L(\mathbf{c})=n$.\footnote{In fact, everywhere below in $M_n(h;\mathbf{c})$ we have $n=L(\mathbf{c})$, making the index $n$ redundant. However, we find beneficial to have it as a ``checksum'' for $\mathbf{c}$.}
We remark that $M_n(h;\mathbf{c})$ does not depend on the order of genes in $P$ but only on their quantity.

The generating function of numbers $M_n(h;\mathbf{c})$ is defined by  
\[
\begin{split}
F(x;u; s_1,s_2,\dots) 
&= \sum_{\mathbf{c}} x^{L(\mathbf{c})-1} \sum_{h=1}^\infty M_{L(\mathbf{c})}(h;\mathbf{c}) u^{h-1}\prod_{i=1}^\infty s_i^{c_i}
\\
&= \sum_{n=1}^\infty x^{n-1} \sum_{\mathbf{c}: L(\mathbf{c}) = n} \sum_{h=1}^\infty M_n(h;\mathbf{c}) u^{h-1}\prod_{i=1}^\infty s_i^{c_i}
\, .
\end{split}
\]
We remark that $F(x;u; s_1,s_2,\dots)$ at $x=0$ equals $s_1$ (which corresponds to $G(P,Q)$, where $Q=P$ consists of $n=1$ gene), while at $u=0$ it enumerates breakpoint graphs $G(P,Q)$ for unichromosomal genomes $Q$.

\begin{theorem}
The following equation, together with the initial condition $F(0;u; s_1,s_2,\dots) = s_1$, uniquely determines the generating function $F(x;u; s_1,s_2,\dots)$. 
\begin{equation}\label{eq:diff_op}
 \begin{split}
\frac{\partial F}{\partial x} &= \sum_{i=2}^\infty \sum_{j=1}^{i-1} (i-1)s_j s_{i-j}\frac{\partial F}{\partial s_{i-1}}  + \sum_{i=2}^\infty (i-1)^2 s_{i}\frac{\partial F}{\partial s_{i-1}} \\
  &+2\sum_{i=2}^\infty \sum_{j=1}^{i-1}j(i-j)s_{i+1}\frac{\partial^2 F}{\partial s_{j}\partial s_{i-j}}+ u\sum_{i=1}^\infty i s_{i+1}\frac{\partial F}{\partial s_{i}}\, .
\end{split}
\end{equation}
 \label{thm:multi}
\end{theorem}

\begin{proof} The theorem statement follows from Lemma~\ref{lem:signed} below, which essentially restates the equation \eqref{eq:diff_op} as equalities of the coefficients of $x^{n-2}u^{h-1}\prod_{i\geq 1} s_i^{c_i}$ in the left- and right-hand sides of \eqref{eq:diff_op}.
Furthermore, these equalities uniquely determine the values of all $M_n(h;\mathbf{c})$ by induction on $n$, thus determining $F(x;u; s_1,s_2,\dots)$.
\end{proof}

\begin{lemma}\label{lem:signed}
For any positive integers $n,h$, we have (the initial condition) 
$$M_1(h;\mathbf{c}) = 
\begin{cases} 1, & \textrm{if } h=1 \textrm{ and } \mathbf{c}=\eb_1; \\
0, & \textrm{otherwise;}
\end{cases}
$$
and for all $n>1$,
\begin{align}
&\hspace{-2em}(n-1) M_n(h;\mathbf{c}) \nonumber \\
  &= \sum_{i=2}^\infty \sum_{j=1}^{i-1} (i-1)(c_{i-1}+1-\delta_{j,1}-\delta_{j, i-1}) M_{n-1}(h;\mathbf{c}+\eb_{i-1}-\eb_{j}-\eb_{i-j}) \label{eq:merge}\\ 
  &+\sum_{i=2}^\infty  (i-1)^2(c_{i-1}+1) M_{n-1}(h;\mathbf{c}+\eb_{i-1}-\eb_{i}) \label{eq:signed_split}\\
  &+ 2\sum_{i=2}^\infty \sum_{j=1}^{i-1} j(i-j)(c_j+1)(c_{i-j}+1+\delta_{j,i-j}) M_{n-1}(h;\mathbf{c}-\eb_{i+1}+\eb_{j}+\eb_{i-j})  \label{eq:signed_merge} \\
  &+ \sum_{i=2}^\infty  (i-1)(c_{i-1}+1) M_{n-1}(h-1;\mathbf{c}+\eb_{i-1}-\eb_{i}) \, , \label{eq:new_chr}
\end{align}
where $\delta_{i,j}$ is the Kronecker delta, and $\eb_i = (\delta_{i,1},\delta_{i,2},\dots)$ is a unit vector (where all coordinates except the $i$-th are zero).\footnote{We remark that in \eqref{eq:merge}-\eqref{eq:new_chr}
all indices of $M$ are in agreement with the corresponding cycle structure, i.e., $L(\mathbf{c})=n$ and each of
$L(\mathbf{c}+\eb_{i-1}-\eb_{j}-\eb_{i-j})$, $L(\mathbf{c}+\eb_{i-1}-\eb_{i})$, $L(\mathbf{c}-\eb_{i+1}+\eb_{j}+\eb_{i-j})$ equals $n-1$.}
\end{lemma}

\begin{proof}
We prove the lemma statement using double counting.\footnote{A similar technique for a different enumeration problem 
was used in~\citet{alexeev2016}.}

Let $n>1$, $Q \in \mathcal{Q}_n(h;\mathbf{c})$, and $l\in\{1,\dots,n-1\}$. We remove gene $l$ from both genomes $P$ and $Q$ to obtain new genomes $P'$ and $Q'$ on $n-1$ genes. 
Then the breakpoint graph $G(P',Q')$ can be obtained from $G(P,Q)$ by removal of vertices $l^t$ and $l^h$ and incident gray edges $\{l^t,a\}$, $\{l^h,c\}$ and 
black edges $\{l^t,b\}$, $\{l^h,d\}$, and addition of a new gray edge $\{a,c\}$ (unless $a=l^h$ and $b=l^t$) and a new black edge $\{b,d\}$ (Fig.~\ref{fig:rem}). 
Clearly, in $G(P,Q)$ vertices $a,b$ belong to the same black-gray cycle and so do vertices $c,d$.
Similarly, in $G(P',Q')$ vertices $b,d$ belong to the same black-gray cycle and so do vertices $a,b$ (if present).

  \begin{figure}[!t]
  \begin{center}
  \includegraphics[width=\textwidth]{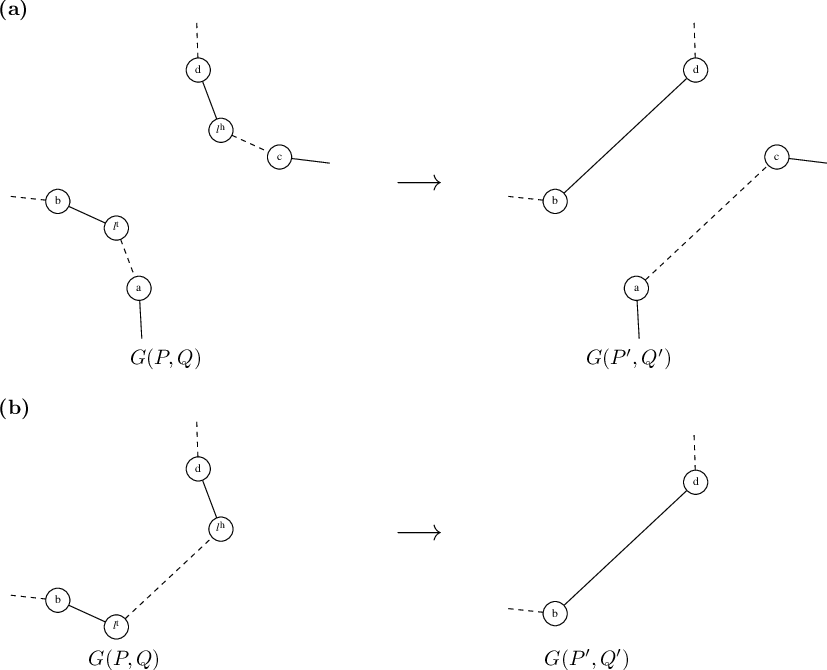}
    \caption{A transformation of breakpoint graphs corresponding to removal of gene $l$ from genomes $P$ and $Q$ resulting in genomes $P'$ and $Q'$.
    \textbf{(a)} The graph $G(P,Q)$ has no gray edge $\{l^t,l^h\}$, i.e., $a \ne l^h$ and $c \ne l^t$.
    \textbf{(b)} The graph $G(P,Q)$ contains the gray edge $\{l^t,l^h\}$, i.e., $a = l^h$ and $c = l^t$.}
    \label{fig:rem}
  \end{center}
  \end{figure}

Below we analyze how the cycle structure of $G(P',Q')$ may differ from the cycle structure of $G(P,Q)$. There are four cases to consider:

\paragraph{Case 1.} Vertices $a, b$ belong to a different cycle in $G(P,Q)$ than vertices $c$ and $d$. 
If these cycles are a $j$-cycle and a $(i-j)$-cycle ($i>j\geq 1$), respectively, 
then $Q' \in \mathcal{Q}_{n-1}(h;\mathbf{c}+\eb_{i-1}-\eb_{j}-\eb_{i-j})$ 
and vertices $a, b, c, d$ belong to the same $(i-1)$-cycle in $G(P',Q')$.

\paragraph{Case 2.} Vertices $a,b,c,d$ belong to the same $(i+1)$-cycle ($i\geq 2$) in $G(P,Q)$ and their order is $(a,l^t,b,\dots,c,l^h,d,\dots)$. 
In this case, $Q' \in \mathcal{Q}_{n-1}(h;\mathbf{c}+\eb_{i}-\eb_{i+1})$ and vertices $a,b,c,d$ belong to the same $i$-cycle in $G(P,Q)$.

\paragraph{Case 3.} Vertices $a,b,c,d$ belong to the same $(i+1)$-cycle ($i\geq 2$) in $G(P,Q)$ and their order is $(a,l^t,b,\dots,d,l^h,c,\dots)$. 
In this case, $Q' \in \mathcal{Q}_{n-1}(h;\mathbf{c}-\eb_{i+1}+\eb_{j}+\eb_{i-j})$, edge $\{a,c\}$ belongs to some $j$-cycle ($1\leq j<i$) in $G(P',Q')$, and edge $\{b,d\}$ belongs to some $(i-j)$-cycle in $G(P',Q')$.

\paragraph{Case 4.} Vertices $a,c$ coincide with $l^h,l^t$, i.e., $a = l^h$ and $c = l^t$.\footnote{We remark that 
a similarly looking case $b=l^h$ and $d=l^t$ is not possible, since $P$ is a unichromosomal genome with $n>1$ genes.} 
This means that gene $l$ forms its own chromosome in $Q$ and this chromosome is removed in $Q'$ (Fig.~\ref{fig:rem}b). 
In this case, vertices $a,b,c,d$ belong to the same $i$-cycle in $G(P,Q)$ for some $i\geq 2$, and $b,d$ belong to an $(i-1)$-cycle in $G(P',Q')$. 
Hence, $Q' \in \mathcal{Q}_{n-1}(h-1;\mathbf{c}+\eb_{i-1}-\eb_{i})$.

We define a function $\Gamma_l$, which maps a genome $Q$ to a pair $(Q',(a,c))$ (Cases 1-3) or a genome $Q'$ (Case 4), where $(a,c)$ is an ordered pair of vertices corresponding to a gray edge in $G(P',Q')$.
For any integers $n>1$ and $l\in\{1,\dots,n-1\}$, we will prove that $\Gamma_l$ is a bijection between 
(i) the $h$-chromosomal genomes $Q$ on $n$ genes; and (ii) the union of the $h$-chromosomal genomes $Q'$ on $n-1$ genes 
with a marked gray edge in $G(P',Q')$ and the $(h-1)$-chromosomal genomes on $n-1$ genes. Namely, we will show that $\Gamma_l$ is invertible. Indeed, given an $h$-chromosomal genome $Q'$ on genes $\{1,2,\dots,n-1\}$ and a pair $(a,c)$, we relabel the genes consecutively into $\{1,2,\dots,l-1,l+1,\dots,n\}$. To reconstruct a genome $Q$ from $Q'$, we insert gene $l$ in between of the genes corresponding to vertices $a$ and $c$ (in the direction from $a$ to $c$). Similarly, given an $(h-1)$-chromosomal genome $Q'$ on genes $\{1,2,\dots,n-1\}$, we relabel its genes and construct genome $Q$ from $Q'$ by adding a new chromosome consisting of a single gene $l$.

To obtain a formula $M_n(h;\mathbf{c})$ for given integer $h\geq 1$ and cycle structure $\mathbf{c}$ (with $n=L(\mathbf{c})$), 
we restrict functions $\Gamma_l$ to the genomes $Q\in\mathcal{Q}_n(h;\mathbf{c})$. Since there are $n-1$ values of $l$, 
the total number of pairs $(Q,\Gamma_l(Q))$ equals $(n-1) M_n(h;\mathbf{c})$. Since each $\Gamma_l$ is a bijection, this amount also equals the sum of
\begin{itemize}
\item 
number of pairs 
$(\Gamma_l^{-1}((Q',(a,c))),(Q',(a,c)))$, where $\Gamma_l^{-1}((Q',(a,c)))\in \mathcal{Q}_n(h;\mathbf{c})$ and $Q'$ belongs to $\mathcal{Q}_{n-1}(h;\mathbf{c}+\eb_{i-1}-\eb_{j}-\eb_{i-j})$, $\mathcal{Q}_{n-1}(h;\mathbf{c}+\eb_{i}-\eb_{i+1})$, or $\mathcal{Q}_{n-1}(h;\mathbf{c}-\eb_{i+1}+\eb_{j}+\eb_{i-j})$ for some $i>j\geq 1$ (Cases 1,2,3, respectively); and
\item
number of pairs $(\Gamma_l^{-1}(Q'),Q')$, where $\Gamma_l^{-1}(Q')\in \mathcal{Q}_n(h;\mathbf{c})$ and $Q'\in \mathcal{Q}_{n-1}(h-1;\mathbf{c}+\eb_{i-1}-\eb_{i})$ (Case 4).
\end{itemize}
We consider Cases 1 and 3 in details.

In Case 1, for any given integers $i>j\geq 1$, we consider a genome $Q' \in \mathcal{Q}_{n-1}(h;\mathbf{c}+\eb_{i-1}-\eb_{j}-\eb_{i-j})$ composed of genes $\{1,2,\dots,n-1\}$ and enumerate the ways to reconstruct some genome $Q \in \mathcal{Q}_n(h;\mathbf{c})$ from $Q'$. 
First, we choose an $(i-1)$-cycle $C$ in $G(P',Q')$, which can be done in $c_{i-1}+1-\delta_{j,1}-\delta_{j, i-1}$ ways. Then we choose an integer $l$ such that the black edge $\{(l-1)^h,l^t\}$ belongs to $C$, which can be done in $i-1$ ways. Then the cycle $C$ has the form $((l-1)^h,l^t,\dots,c,a,\dots)$, where there are $2i-2j$ edges between vertices $l^t$ and $c$ (and thus $\{a,c\}$ represents a gray edge in $C$). Then we reconstruct a genome $Q$ as $Q=\Gamma_l^{-1}((Q',(a,c)))$. Summing over the values of $i,j$ gives the term \eqref{eq:merge} for the total number of such genomes $Q$.

In Case 3, for any given integers $i>j\geq 1$, we consider a genome $Q' \in \mathcal{Q}_{n-1}(h;\mathbf{c}-\eb_{i+1}+\eb_{j}+\eb_{i-j})$ composed of genes $\{1,2,\dots,n-1\}$ and enumerate the number of ways to reconstruct some genome $Q \in \mathcal{Q}_n(h;\mathbf{c})$ from $Q'$. First, we choose a $j$-cycle and an $(i-j)$-cycle in $G(P',Q')$, which can be done in $(c_j+1)(c_{i-j}+1+\delta_{j,i-j})$ ways. Then we choose a gray edge $\{u,v\}$ in the $j$-cycle (in $j$ ways) and choose an integer $l$ such that the black edge $\{(l-1)^h,l^t\}$ is in the $(i-j)$-cycle (in $i-j$ ways). Then we reconstruct a genome $Q$ in two ways: $Q=\Gamma_l^{-1}((Q',(u,v)))$ and $Q=\Gamma_l^{-1}((Q',(v,u)))$, which gives factor $2$. Summing over the values of $i,j$ gives the term \eqref{eq:signed_merge} for the total number of such genomes $Q$.

Cases 2 and 4 follow similarly and deliver the terms \eqref{eq:signed_split} and \eqref{eq:new_chr}, respectively. 
\end{proof}

\section{Applications}

\subsection{Hultman Numbers}
Let $P$ be a fixed linear unichromosomal genome on $n$ co-oriented genes and $H(n,n+1-d)$ be the number of 
linear unichromosomal genomes $Q$ on the same co-oriented genes such that the $2$-break distance between $P$ and $Q$ is $d$.
The numbers $H(n,m)$ are called \emph{Hultman numbers}~\citep{doignon2007,bona2009,alexeev2014} and present in the OEIS~\citep{oeis} as the sequence \texttt{A164652}.
The problem  of enumerating linear unichromosomal genomes can be reduced to enumerating circular genomes as follows. One can add a virtual gene $0$ to the genomes $P$ and $Q$ in between of the first and last genes on their chromosomes, making them circular. Then the 2-break distance between $P$ and $Q$ equals $n+1-m$, where $m$ is the number of cycles in the (modified) breakpoint graph $G(P,Q)$.

The Hultman numbers can be obtained from a modification of Theorem \ref{thm:multi}. Namely, let  $P$ be a fixed unichromosomal circular genome with genes $\{1,2,\dots,n\}$ and let $\mathcal{Q^+}_n(h;\mathbf{c})$ be the set of $h$-chromosomal circular genomes $Q$ on the same co-oriented $n$ genes such that $G(P,Q)$ has cycle structure $\mathbf{c}$. Denote
the cardinality of $\mathcal{Q^+}_n(h;\mathbf{c})$ by $M^+_n(h;\mathbf{c})$.

The generating functions of numbers $M^+_n(h;\mathbf{c})$ is defined by  
\[
G(x;u;s_1,s_2,\dots) 
=\sum_{n=1}^\infty x^{n-1}  \sum_{h=1}^\infty u^{h-1} \sum_{\mathbf{c}: L(\mathbf{c}) = n} M^+_n(h;\mathbf{c})\prod_{i=1}^\infty s_i^{c_i}
\, .
\]

\begin{theorem}\label{thm:hultman}
The following equation, together with the initial condition $G(0;u;s_1,s_2,\dots) = s_1$, uniquely determines the generating function $G(x;u;s_1,s_2,\dots)$. 
\[
\begin{split}
  \frac{\partial G}{\partial x} &= \sum_{i=2}^\infty \sum_{j=1}^{i-1} (i-1)s_j s_{i-j}\frac{\partial G}{\partial s_{i-1}}\\
  &+\sum_{i=2}^\infty \sum_{j=1}^{i-1}j(i-j)s_{i+1}\frac{\partial^2 G}{\partial s_{j}\partial s_{i-j}}\\
  &+u\sum_{i=1}^\infty i s_{i+1}\frac{\partial G}{\partial s_{i}}\, .
\end{split}
\]
 
\end{theorem}

\begin{proof}
The proof is similar to the proof of Theorem~\ref{thm:multi} and Lemma \ref{lem:signed}, except that genome $Q$ here has to have co-oriented genes and
thus there is no Case 2 and there is no factor $2$ for Case 3.
\end{proof}

Let $F_n(u; s_1,s_2,\dots)$ and $G_n(u;s_1,s_2,\dots)$ be the coefficients of $x^{n-1}$ in $F(x;u; s_1,s_2,\dots)$ and 
$G(x;u; s_1,s_2,\dots)$, respectively. 
The first few values\footnote{These values are computed with Mathematica code given in Appendix.} of $F_n(0;s_1,s_2,\dots)$ and $G_n(0;s_1,s_2,\dots)$ corresponding to unichromosomal genomes are listed below:
\begin{align*}
F_1(0;s_1,s_1,\dots) &= s_1,\\ 
F_2(0;s_1,s_2,\dots) &= s_1^2+s_2,\\
F_3(0;s_1,s_2,\dots) &= s_1^3+3s_1s_2+ 4s_3,\\
F_4(0;s_1,s_2,\dots) &= s_1^4+6s_1^2s_2 + (5s_2^2+16s_1s_3)+20s_4,\\
F_5(0;s_1,s_2,\dots) &= s_1^5+10s_1^3s_2+(40s_1^2s_3 + 25s_1s_2^2)+(100s_1s_4+60s_2s_3) + 148s_5.\\
\end{align*}
\begin{align*}
 G_1(0;s_1,s_1,\dots) &= s_1,\\
 G_2(0;s_1,s_2,\dots) &= s_1^2,\\
 G_3(0;s_1,s_2,\dots) &= s_1^3+ s_3,\\
 G_4(0;s_1,s_2,\dots) &= s_1^4+(4s_1s_3 + s_2^2),\\
 G_5(0;s_1,s_2,\dots) &= s_1^5+(10s_1^2s_3 + 5s_1s_2^2) + 8s_5,\\
 G_6(0;s_1,s_2,\dots) &= s_1^6+(20s_1^3s_3 + 15s_1^2s_2^2) + (48s_1s_5+ 12s_3^2+24s_2s_4).\\
 \end{align*}

Taking $s_i = s$ for all $i=1,2,\dots$, we get
$$G_n(0;s,s,\dots) = \sum_{m=1}^{n+1} H(n-1,m)s^m \, .$$
In particular, we obtain the following formula for Hultman numbers:
$$H(n-1,m) = \sum_{\mathbf{c} \in \mathcal{C}_{n,m}} M^+_n(1;\mathbf{c}) \, ,$$
where $\mathcal{C}_{n,m} = \{\mathbf{c}:L(\mathbf{c}) = n\textrm{ and } \sum_{i=1}^{n} c_i = m\}$.

\citet{signed2013} introduced the problem of enumerating linear unichromosomal genomes, where genes may be arbitrarily oriented. 
The corresponding \emph{signed Hultman numbers} $H^{\pm}(n,m)$ form the sequence \texttt{A189507} in the OEIS.
Theorem~\ref{thm:multi} allows us to compute these numbers as follows:
\begin{equation}
H^\pm(n-1,m) = \sum_{\mathbf{c} \in \mathcal{C}_{n,m}} M_n(1;\mathbf{c}) \, .
\label{eq:HMrel}
\end{equation}
The first few numbers $H(n,m)$ and $H^\pm(n,m)$ are listed in Table~\ref{tab:hultman}.

\begin{table}[!htb]
\caption{Values of Hultman numbers.}\label{tab:hultman}
\begin{subtable}[t]{0.48\textwidth}
\caption{Values of $H(n,m)$.}
\label{tab:hult2}
\centering
\begin{tabular}{| c | r r r r  r r | }
\hline
  $n\backslash m$ &   $1$ &    $2$ & $3$ & $4$ & $5$ & $6$\\
  \hline
  0 & 1 &  &  & & & \\
  1 & 0 & 1 & & & &\\
  2 & 1 & 0 & 1 & & &\\
  3 & 0 & 5 & 0 & 1 & &\\
  4 & 8 & 0 & 15 & 0 & 1 & \\
  5 & 0 & 84 & 0 & 35 & 0 & 1\\
  \hline
\end{tabular}
\end{subtable}
~
\begin{subtable}[t]{0.48\textwidth}
\caption{Values of $H^\pm(n,m)$.}
\label{tab:shult2}
\centering
\begin{tabular}{| c | r r r r r r| }
\hline
  $n\backslash m$ &   $1$ &    $2$ & $3$ & $4$ & $5$ & $6$ \\
  \hline
  0 & 1 & & & & &\\
  1 & 1 & 1 & & & & \\
  2 & 4 & 3 & 1& & &\\
  3 & 20 & 21 & 6 & 1 & &\\
  4 & 148 & 160 & 65 & 10 &  1&\\
  5 & 1348 & 1620 & 701 & 155 & 15 & 1\\
  \hline
\end{tabular}
\end{subtable}
\end{table}

\subsection{Bell Polynomials}
The numbers $M_n(h;\mathbf{c})$ have multiple connections to well-known combinatorial objects. 
Some of these connections are straightforward, and some appear to be new. 

It is easy to see that in the unichromosomal case, $G_n(0;1,1,\dots)$ enumerates permutations of order $n-1$, and so
$$G_n(0;1,1,\dots) = (n-1)! \, .$$
Similarly, $F_n(0;1,1,\dots)$ enumerates signed permutations of order $n-1$, and so
$$F_n(0;1,1,\dots) = 2^{n-1} (n-1)! \, .$$
In the multichromosomal case, we get more general formulae:
$$G_n(u;1,1,\dots) = \sum_{h=1}^n \left[n\atop h\right] u^{h-1} \,$$
and
$$F_n(u;1,1,\dots) = \sum_{h=1}^n 2^{n-h} \left[n\atop h\right] u^{h-1} \, ,$$
where $\left[n\atop h\right]$ are unsigned Stirling numbers of the first kind (\texttt{A094638} in the OEIS).
Moreover, for $u=1$, we have
\begin{equation}\label{eq:Gn1}
G_n(1;s_1,s_2,\dots) = \sum_{\mathbf{c}: L(\mathbf{c}) = n} \frac{n!}{\prod_{i=1}^n c_i!} \prod_{i=1}^n \left(\frac{s_i}{i}\right)^{c_i} \, .
\end{equation}
The numbers $L(\mathbf{c})!/(c_1! 1^{c_1}c_2! 2^{c_2}\dots)$ 
enumerate permutations with the cycle structure $\mathbf{c}$ and form the sequence \texttt{A124795} in the OEIS.
The functions $G_n(1;s_1,s_2,\dots)$ are closely related to the complete exponential Bell polynomials~\citep[Section~3.3]{Comtet1974}
\begin{equation}\label{eq:Bn}
Y_n(x_1,x_2,\dots) = \sum_{\mathbf{c}: L(\mathbf{c}) = n} \frac{n!}{\prod_{i=1}^n c_i!} \prod_{i=1}^n \left(\frac{x_i}{i!}\right)^{c_i} \, .
\end{equation}
Namely, from \eqref{eq:Gn1} and \eqref{eq:Bn} it follows that
$$G_n\left(1;\frac{s_1}{0!},\frac{s_2}{1!},\frac{s_3}{2!},\dots,\frac{s_k}{(k-1)!},\dots\right) = Y_n(s_1,s_2,\dots) \, . $$
Hence, Theorem~\ref{thm:hultman} implies the following (apparently new) differential equation for Bell polynomials:
 \begin{align*}
  (n-1) Y_n (x_1,x_2,\dots) &= \sum_{i=2}^\infty \sum_{j=1}^{i-1} (i-1)\binom{i-2}{j-1} x_j x_{i-j}\frac{\partial Y_{n-1}}{\partial x_{i-1}}\\
  &+\sum_{i=2}^\infty \sum_{j=1}^{i-1} \frac{x_{i+1}}{\binom{i}{j}}\frac{\partial^2 Y_{n-1}}{\partial x_{j}\partial x_{i-j}}\\
  &+\sum_{i=2}^\infty x_{i}\frac{\partial Y_{n-1}}{\partial x_{i-1}}\, .
  \end{align*}
  
\subsection{Distribution Of $k$-Break Distances}

Let $H^h_k(n,d)$ be the number of $h$-chromosomal circular
genomes with $n$ genes at the $k$-break distance $d$ from a fixed unichromosomal circular  genome.
For $k=2$ and $h=1$, these numbers represent signed Hultman numbers: $H^1_2(n,d)=H^\pm(n-1,n-d)$.

Using formulae~\eqref{eq:d3} and \eqref{eq:d4}, we can further obtain $H^h_3(n,d)$ and $H^h_4(n,d)$. 
The first few numbers $H^1_3(n,d)$, $H^2_3(n,d)$, $H^1_4(n,d)$, and $H^2_4(n,d)$ are listed in Table~\ref{tab:h3}.

\begin{table}[!htb]
\caption{Values of generalized Hultman numbers.}\label{tab:h3}
\begin{subtable}{0.48\textwidth}
\caption{Values of $H^1_3(n,d)$.}
\centering
\begin{tabular}{| c | r r r r | }
\hline
  $n\backslash d$ & $0$ &   $1$ &    $2$ & $3$ \\
  \hline
  1 & 1 & 0 & 0 & 0\\
  2 & 1 & 1 &  0&  0 \\
  3 & 1 & 7 & 0& 0 \\
  4 & 1 & 22 & 25 & 0 \\
  5 & 1 & 50 & 333 & 0 \\
  6 & 1 & 95 & 1851 & 1893 \\
  7 & 1 & 161 & 6839 & 39079 \\
  \hline
\end{tabular}
\end{subtable}
~
\begin{subtable}{0.48\textwidth}
\caption{Values of $H^2_3(n,d)$.}
\centering
\begin{tabular}{| c | r r r | }
\hline
  $n\backslash d$ & $1$ &    $2$ & $3$ \\
  \hline
  1 & 0&  0&  0 \\
  2 & 1 &  0& 0  \\
  3 & 6 & 0 &0 \\
  4 & 18 & 26 & 0 \\
  5 & 40 & 360 &0 \\
  6 & 75 & 2034 & 2275\\
  7 & 126 & 7588 & 48734\\
  \hline
\end{tabular}
\end{subtable}
\\~~~\\

\begin{subtable}[h]{0.48\textwidth}
\caption{Values of $H^1_4(n,d)$.}
\centering
\begin{tabular}{| c | r r r r | }
\hline
  $n\backslash d$ & $0$ &   $1$ &    $2$ & $3$  \\
  \hline
  1 & 1 & 0  & 0 & 0 \\
  2 & 1 & 1 & 0 & 0  \\
  3 & 1 & 7 & 0 & 0 \\
  4 & 1 & 47 & 0 & 0 \\
  5 & 1 & 175 & 208 & 0  \\
  6 & 1 & 470 & 3369 & 0  \\
  7 & 1 & 1036 & 45043 & 0  \\
  8 & 1 & 2002 & 315213 & 327904  \\
  \hline
\end{tabular}
\end{subtable}
~
\begin{subtable}[h]{0.48\textwidth}
\caption{Values of $H^2_4(n,d)$.}
\centering
\begin{tabular}{| c |  r  r  r | }
\hline
  $n\backslash d$ &    $1$ & $2$ & $3$\\
  \hline
  1 & 0 & 0 & 0    \\
  2 & 1 & 0 & 0    \\
  3 & 6 & 0 & 0  \\
  4 & 44 & 0 & 0   \\
  5 & 170 & 230 & 0   \\
  6 & 465 & 3919 & 0   \\
  7 & 1036 & 55412 & 0\\
  8 & 2016 & 396764 & 437572\\ 
  \hline
\end{tabular}
\end{subtable}
\end{table}

\subsection{Sampling Of Random Genomes}
Theorem~\ref{thm:multi} and Lemma~\ref{lem:signed} allow us to sample a (uniformly) random genome $Q$ with given number of genes $n$, number of chromosomes $h$, and cycle structure $\mathbf{c}$ of the breakpoint graph $G(P,Q)$. 
Namely, we define a Markov chain $\mathcal{M}$ as follows:
\begin{itemize}
 \item the states of $\mathcal{M}$ are genome classes $\mathcal{Q}_{n}(h;\mathbf{c})$;
 \item the probability of transition between $\mathcal{Q}_{n}(h;\mathbf{c})$ and $\mathcal{Q}_{n-1}(h;\mathbf{c}+\eb_{i-1}-\eb_{j}-\eb_{i-j})$ (for any $i\geq 2$ and $1\leq j < i$) is 
 $$ \frac{(i-1)(c_{i-1}+1-\delta_{j,1}-\delta_{j, i-1}) M_{n-1}(h;\mathbf{c}+\eb_{i-1}-\eb_{j}-\eb_{i-j})}{(n-1) M_n(h;\mathbf{c})} \, ;$$
 \item the probability of transition between $\mathcal{Q}_{n}(h;\mathbf{c})$ and $\mathcal{Q}_{n-1}(h;\mathbf{c}+\eb_{i-1}-\eb_{i})$ (for any $i \geq 2$) is 
 $$ \frac{(i-1)^2(c_{i-1}+1) M_{n-1}(h;\mathbf{c}+\eb_{i-1}-\eb_{i})}{(n-1) M_n(h;\mathbf{c})} \, ;$$
 \item the probability of transition between $\mathcal{Q}_{n}(h;\mathbf{c})$ and $\mathcal{Q}_{n-1}(h;\mathbf{c}-\eb_{i+1}+\eb_{j}+\eb_{i-j})$ (for any $i \geq 2$ and $1\leq j < i$) is 
 $$ \frac{2j(i-j)(c_j+1)(c_{i-j}+1+\delta_{j,i-j}) M_{n-1}(h;\mathbf{c}-\eb_{i+1}+\eb_{j}+\eb_{i-j})}{(n-1) M_n(h;\mathbf{c})} \, ;$$
\item the probability of transition between $\mathcal{Q}_{n}(h;\mathbf{c})$ and $\mathcal{Q}_{n-1}(h-1;\mathbf{c}+\eb_{i-1}-\eb_{i})$ (for any $i\geq 2$) is 
 $$ \frac{(i-1)(c_{i-1}+1) M_{n-1}(h-1;\mathbf{c}+\eb_{i-1}-\eb_{i})}{(n-1) M_n(h;\mathbf{c})} \, ;$$
  \item the probability of transition between $\mathcal{Q}_{1}(1;\eb_1)$ and itself is equal to $1$; 
 \item in the other cases, the transition probability equals to $0$.
\end{itemize}
Lemma \ref{lem:signed} implies that the Markov chain $\mathcal{M}$ is well-defined. 
For any initial state $\mathcal{Q}_{n}(h;\mathbf{c})$, the process after $n-1$ steps comes into 
the terminal state $\mathcal{Q}_{1}(1;\eb_1)$, which consists of a single genome. 

To sample a random genome  $Q \in \mathcal{Q}_{n}(h;\mathbf{c})$, 
we first sample a random path $(\mathcal{Q}_{n}, \mathcal{Q}_{n-1},\dots,\mathcal{Q}_1)$ 
starting at $\mathcal{Q}_{n} = \mathcal{Q}_{n}(h;\mathbf{c})$ and ending at the termination state $\mathcal{Q}_1=\mathcal{Q}_{1}(1;\eb_1)$. 
We start with $Q\in \mathcal{Q}_1$ (i.e., $Q$ is a genome with a single gene) and for every $j$ from $1$ to $n-1$,
we randomly add a gene into $Q$ such that the resulting genome belongs to $\mathcal{Q}_{j+1}$. 
By construction, at the end of this process the genome $Q$ represents a uniformly random element of $\mathcal{Q}_{n}(h;\mathbf{c})$.

\section{Discussion}
In the current work, we address the problem of enumeration of genomes with $n$ genes that are at a given $k$-break distance from a fixed unichromosomal genome.
It is known that the $k$-break distance between two genomes can be computed in terms of cycle lengths in the breakpoint graph of these genomes~\citep{alekseyev2008}.

Our main result is the recurrent formula for the numbers $M_n(h;\mathbf{c})$ (and their generating function) of breakpoint graphs with the cycle structure $\mathbf{c}$
of $h$-chromosomal genomes with $n$ genes.
We show connection between these numbers and various combinatorial objects (such as Bell polynomials)
and further compute numbers $H^h_k(n,d)$ of $h$-chromosomal genomes with $n$ genes at the $k$-break distance $d$ from a fixed unichromosomal genome,
which generalize Hultman numbers~\citep{hultman1999,doignon2007,bona2009,alexeev2014,signed2013}.
  
We believe that our approach can further lead to finding a formula for the numbers $H^h_k(n,d)$ and then to evaluating the asymptotic distribution of the $k$-break distances for a general $k$.
Other open questions of interest include enumeration of genomes $Q$ at a given $k$-break distance from a fixed genome $P$,
where (i) $P$ is unichromosomal and $Q$ is linear multichromosomal (the case $k=2$ was addressed by \citet{multi2014}); or (ii) $P$ and $Q$ are both multichromosomal.
Both questions may be addressed under the assumption of co-oriented or arbitrarily oriented genes.
Defining \emph{proper} $k$-breaks as those that are not $(k-1)$-breaks, we may ask similar questions for the \emph{graded} $(2,3,\dots,k)$-break distance specifying the number of proper $i$-breaks for each $i=2,3,
\dots,k$.
Further assuming that proper $k$-breaks for different $k$ have different rates in the course of evolution, we may be able to estimate these rates from given (extant) genomes,
using the technique proposed by \citet{alexeev2015transposition} for $k=2,3$.

\section*{Acknowledgements}
The work is supported by the National Science Foundation under the grant No. IIS-1462107. 
 
\bibliographystyle{apalike}
\bibliography{hultman.bib}

\begin{thebibliography}{}

\bibitem[Alekseyev and Pevzner, 2008]{alekseyev2008}
Alekseyev, M. and Pevzner, P. (2008).
\newblock Multi-break rearrangements and chromosomal evolution.
\newblock {\em Theoretical Computer Science}, 395(2):193--202.

\bibitem[Alexeev et~al., 2015]{alexeev2015transposition}
Alexeev, N., Aidagulov, R., and Alekseyev, M.~A. (2015).
\newblock A computational method for the rate estimation of evolutionary
  transpositions.
\newblock In Ortu\~no, F. and Rojas, I., editors, {\em Proceedings of the 3rd
  International Work-Conference on Bioinformatics and Biomedical Engineering
  (IWBBIO)}, volume 9043 of {\em Lecture Notes in Computer Science}, pages
  471--480.

\bibitem[Alexeev et~al., 2016]{alexeev2016}
Alexeev, N., Andersen, J., Penner, R., and Zograf, P. (2016).
\newblock Enumeration of chord diagrams on many intervals and their
  non-orientable analogs.
\newblock {\em Advances in Mathematics}, 289:1056 -- 1081.

\bibitem[Alexeev and Zograf, 2014]{alexeev2014}
Alexeev, N. and Zograf, P. (2014).
\newblock Random matrix approach to the distribution of genomic distance.
\newblock {\em Journal of Computational Biology}, 21(8):622--631.

\bibitem[B{\'o}na and Flynn, 2009]{bona2009}
B{\'o}na, M. and Flynn, R. (2009).
\newblock The average number of block interchanges needed to sort a permutation
  and a recent result of stanley.
\newblock {\em Information Processing Letters}, 109(16):927--931.

\bibitem[Caprara, 1997]{Caprara1997}
Caprara, A. (1997).
\newblock Sorting by reversals is difficult.
\newblock In {\em Proceedings of the first annual international conference on
  Computational molecular biology (RECOMB)}, pages 75--83.

\bibitem[Comtet, 1974]{Comtet1974}
Comtet, L. (1974).
\newblock {\em {Advanced Combinatorics}}.
\newblock D. Reidel Publishing Company, Dordrecht, Holland.

\bibitem[Doignon and Labarre, 2007]{doignon2007}
Doignon, J.-P. and Labarre, A. (2007).
\newblock {On Hultman numbers}.
\newblock {\em Journal of Integer Sequences}, 10(6):Article 07.6.2.

\bibitem[Feij{\~a}o et~al., 2014]{multi2014}
Feij{\~a}o, P., Martinez, F.~V., and Th{\'e}venin, A. (2014).
\newblock {On the Multichromosomal Hultman Number}.
\newblock In Campos, S., editor, {\em Proceedings of the 9th Brazilian
  Symposium on Bioinformatics (BSB)}, volume 8826 of {\em Lecture Notes in
  Computer Science}, pages 9--16.

\bibitem[Grusea and Labarre, 2013]{signed2013}
Grusea, S. and Labarre, A. (2013).
\newblock The distribution of cycles in breakpoint graphs of signed
  permutations.
\newblock {\em Discrete Applied Mathematics}, 161(10):1448--1466.

\bibitem[Hultman, 1999]{hultman1999}
Hultman, A. (1999).
\newblock Toric permutations.
\newblock {\em Master's thesis, Dept. of Mathematics, KTH, Stockholm, Sweden}.

\bibitem[Stephens et~al., 2011]{chromo2011}
Stephens, P.~J., Greenman, C.~D., Fu, B., Yang, F., Bignell, G.~R., Mudie,
  L.~J., Pleasance, E.~D., Lau, K.~W., Beare, D., Stebbings, L.~A., et~al.
  (2011).
\newblock Massive genomic rearrangement acquired in a single catastrophic event
  during cancer development.
\newblock {\em Cell}, 144(1):27--40.

\bibitem[{The OEIS Foundation}, 2016]{oeis}
{The OEIS Foundation} (2016).
\newblock {\em {The On-Line Encyclopedia of Integer Sequences}}.
\newblock Published electronically at \url{http://oeis.org}.

\bibitem[Weinreb et~al., 2014]{weinreb2014}
Weinreb, C., Oesper, L., and Raphael, B.~J. (2014).
\newblock Open adjacencies and $k$-breaks: detecting simultaneous
  rearrangements in cancer genomes.
\newblock {\em BMC Genomics}, 15(Suppl 6):S4.

\bibitem[Yancopoulos et~al., 2005]{yancopoulos2005}
Yancopoulos, S., Attie, O., and Friedberg, R. (2005).
\newblock Efficient sorting of genomic permutations by translocation, inversion
  and block interchange.
\newblock {\em Bioinformatics}, 21(16):3340--3346.

\end{thebibliography}


\clearpage
\newpage

\section*{Appendix. Mathematica Code}
Here we provide Wolfram Mathematica code for computing the functions $G_0(u;s_1,s_2,\dots)$, $G_n(u;s_1,s_2,\dots)$, $F_n(0;s_1,s_2,\dots)$, $F_n(u;s_1,s_2,\dots)$:
\begin{verbatim}
(*Implementation of the summands in the formula in Theorem 3.1*)
L0[f_, n_] := 
 Sum[Sum[(i - 1)*s[j]*s[i - j]*D[f, s[i - 1]], {j, 1, i - 1}], {i, 2, 
   n}]
L1[f_, n_] := Sum[(i - 1)^2*s[i]*D[f, s[i - 1]], {i, 2, n}]
L2[f_, n_] := 
  Sum[s[i + 1]*Sum[j*(i - j)*D[f, s[j], s[i - j]], {j, 1, i - 1}], {i,
     2, n}];
Ln[f_, n_] := Sum[(i - 1)*u*s[i]*D[f, s[i - 1]], {i, 2, n}];
FG[n_, orient_, multichr_] := {ff := {s[1]}; Do[g := Last[ff];
   f := 1/k*(L0[g, n] + orient*L1[g, n] + (1 + orient)*L2[g, n] 
      + multichr*Ln[g, n]);
   AppendTo[ff, Simplify[f]], {k, n}];
  ff[[n]]}

(*Implementation of function G_n(0;s1,s2,...)*)
G0[n_] := FG[n, 0, 0]
(*Implementation of function G_n(u;s1,s2,...)*)
Gu[n_] := FG[n, 0, 1]
(*Implementation of function F_n(0;s1,s2,...)*)
F0[n_] := FG[n, 1, 0]
(*Implementation of function F_n(u;s1,s2,...)*)
Fu[n_] := FG[n, 1, 1]
\end{verbatim}

\end{document}